\documentstyle[epsfig]{elsart}

\begin{document}
\begin{frontmatter}
\title{Interference of Conversion and Bremsstrahlung Amplitudes in the Decay
$K_L \to  \mu^+ \mu^- \gamma$}

\author{P.~Poulose and L.~M.~Sehgal}
\address{Institute of Theoretical Physics E, RWTH Aachen, D-52056 Aachen, Germany}

\begin{abstract}
In the region of large $\mu^+ \mu^-$ invariant mass, the decay spectrum of $K_L \to \mu^+ \mu^- \gamma$
deviates from the Dalitz pair spectrum, as a result of interference between conversion
($K_L \to \gamma^* \gamma \to \mu^+ \mu^- \gamma$) and bremsstrahlung amplitudes. The latter is proportional
to the $K_L \to  \mu^+ \mu^-$ matrix element, whose $2 \gamma$-absorptive part appears to dominate the
observed $K_L \to  \mu^+ \mu^-$ decay rate. We examine the extent to which a scrutiny of the
$K_L \to  \mu^+ \mu^- \gamma$ spectrum in the end-point region could provide evidence on the real part of
the $K_L \to  \mu^+ \mu^-$ amplitude. As a by-product, we obtain the absorptive part of the
$K_L \to  \gamma^* \gamma$ form factor, using data on the $K_L \to  \pi^+ \pi^- \gamma$ spectrum.
\end{abstract}
\end{frontmatter}

It is customary to interpret the decays $K_L \to  e^+ e^- \gamma$ and $K_L \to  \mu^+ \mu^- \gamma$ in terms
of a Dalitz pair process $K_L \to \gamma^* \gamma \to  l^+ l^- \gamma$. The branching ratio and the lepton
mass spectrum are fitted to an $s$-dependent $K_L \gamma^* \gamma$ vertex
$f_{K \gamma \gamma}(s) = f_{K \gamma \gamma}(0) f(s)$, where $f_{K \gamma \gamma}(0)$ is a dimensionless
parameter related to the decay width of $K_L \to \gamma \gamma$ by
\begin{equation}
\left| f_{K \gamma \gamma}(0) \right|^2 = \frac{64 \pi}{m_K} \Gamma (K_L \to \gamma \gamma)
\end{equation}
and $f(s)$ is a form factor. The spectrum in the invariant mass of the lepton pair ($s = (p_+ + p_-)^2$) is
given by~\cite{Sehgal73}
\begin{eqnarray}
\frac{d \Gamma (K_L \to l^+ l^- \gamma ) / ds}{\Gamma (K_L \to \gamma \gamma)} & = & \frac{2 \alpha}{3 \pi}
\left( 1 - \frac{s}{m_K^2} \right)^3 \left( 1 + \frac{2 m^2_l}{s} \right) \left( 1 - \frac{4 m^2_l}{s}
\right)^{1/2} \nonumber \\
& & \times \frac{1}{s} \left| f(s) \right|^2 \label{Dalitzspec}
\end{eqnarray}

A popular choice of the form factor $f(s)$ is the BMS parametrization~\cite{Bergstroemetal}
\begin{equation}
f_{BMS}(s) = \frac{1}{1- \frac{s}{m^2_{\rho}}} + \frac{2.3 \alpha_{K^*}}{1 - \frac{s}{m^2_{K^*}}}
\left[ \frac{4}{3} - \frac{1}{1- \frac{s}{m^2_{\rho}}} - \frac{1}{9} \frac{1}{1- \frac{s}{m^2_{\omega}}} -
\frac{2}{9} \frac{1}{1- \frac{s}{m^2_{\phi}}} \right] \label{BMSPara}
\end{equation}
where the first term is the $\rho$-dominance approximation, and the remainder is a correction term
depending on an unknown parameter $\alpha_{K^*}$. Measurements on the decays
$K_L \to e^+ e^- \gamma$~\cite{Fanti;LaDue} and
$K_L \to  \mu^+ \mu^- \gamma$~\cite{AlaviHarati;Fanti,BreeseQuinn} can be understood in a fairly
consistent way with a value $\alpha_{K^*} \approx - 0.19$. Other one-parameter forms for $f(s)$ have also
been discussed in the literature~\cite{DAmbrosioetal}.

In the end-point region $s \approx m^2_K$, one expects the decay $K_L \to l^+ l^- \gamma$ to show a deviation
from the Dalitz pair spectrum Eq. (\ref{Dalitzspec}), due to internal bremsstrahlung from the underlying
transition $K_L \to l^+ l^-$. Because of the chiral suppression of the latter, the bremsstrahlung effect is
of relevance mainly for the channel $K_L \to \mu^+ \mu^- \gamma$. Considering that the branching ratio
$Br(K_L \to \mu^+ \mu^-)$ is $\approx 7 \times 10^{-9}$, one expects the bremsstrahlung contribution to
$Br(K_L \to \mu^+ \mu^- \gamma)$ to be of order $10^{-11}$, and confined to photons of very low energy.
There is, however, the possibility of an interference effect between the bremsstrahlung and Dalitz pair
amplitudes that could conceivably probe the real part of the $K_L \to \mu^+ \mu^-$ amplitude. It is this
possibility that we wish to explore in this paper.

It may be recalled that if the $K_L \to \mu^+ \mu^-$ amplitude is parametrized as
$f_{Kll} \overline{u} \gamma_5 v$, the decay rate is
\begin{eqnarray}
\Gamma (K_L \to l^+ l^-) & = & \left| f_{kll} \right|^2 \frac{m_K v_0}{8 \pi}, \nonumber \\
v_0 & = & \left( 1 - \frac{4 m_l^2}{m_K^2} \right)^{1/2}
\end{eqnarray}
The imaginary part of the $K_L \to l^+ l^-$ amplitude can be reliably calculated in terms of
$K_L \to 2 \gamma$, by considering the absorptive contribution of the two-photon intermediate state
($K_L \to \gamma \gamma \to l^+ l^-$), with the result~\cite{Sehgal69}
\begin{equation}
\left| {\rm Im} f_{Kll} \right| = \frac{\alpha}{4 v_0} \frac{m_l}{m_K} \ln \left( \frac{1+ v_0}{1- v_0} \right)
\left| f_{K \gamma \gamma} (0) \right|
\end{equation}
The measured branching ratio of $K_L \to \mu^+ \mu^-$ is almost saturated by the two-photon absorptive part,
the real part being limited by~\cite{AlaviHarati87}
\begin{equation}
\left| {\rm Re} f_{Kll} / {\rm Im} f_{Kll} \right| \leq 0.23 \, \, (90 \% \, {\rm C.L.}) \label{KTeVMeasure}
\end{equation}

\begin{figure}
\center
\makebox[6.5cm]{
\resizebox{6.5cm}{4.9cm}
{\includegraphics*{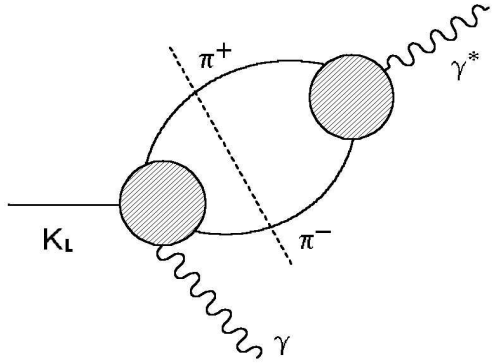}}
}
\caption{Model for imaginary part of $K_L \gamma^* \gamma$ vertex.\label{FeynGraph}}
\end{figure}

In considering the possible interference of bremsstrahlung and conversion amplitudes, it is necessary
to take account of a possible imaginary part in the $K_L \gamma^* \gamma$ form factor $f(s)$. Such an
imaginary part is expected, quite generally, in the region $s > 4 m^2_{\pi}$, and is not included in the
BMS parametrization Eq. (\ref{BMSPara}). As a simple model for ${\rm Im} f(s)$, we consider a $\pi^+ \pi^-$
intermediate state in the virtual photon channel. One can then use unitarity to obtain ${\rm Im} f(s)$ in
terms of the form factor characterising the $K_L \to \pi^+ \pi^- \gamma$ vertex, and the electromagnetic
form factor of the pion (Fig.~\ref{FeynGraph}). We define the invariant amplitude for the direct emission
($M1$) transition $K_L \to \pi^+ \pi^- \gamma$ in the conventional way~\cite{Sehgal;Leusen}
\begin{equation}
{\cal M} \left( K_L \to \pi^+(p_+) \pi^-(p_-) \gamma (\epsilon, k) \right) = \epsilon_{\mu \nu \rho \sigma}
\epsilon ^{* \mu} k^{\nu} p_+^{\rho} p_-^{\sigma} C g_{M1}(s)
\end{equation}
where $s= (p_+ + p_-)^2$ and $C$ is a normalization factor given by
\begin{eqnarray}
C & = & e \frac{|f_S|}{m_K^4} \nonumber \\
|f_S| & = & \left[ \frac{16 \pi m_K}{\beta_0} \Gamma (K_S \to \pi^+ \pi^- ) \right]^{1/2} \\
\beta_0 & = & \left( 1 - \frac{4 m^2_{\pi}}{m^2_K} \right)^{1/2} \nonumber
\end{eqnarray}
Experiments have determined the form factor $g_{M1}(s)$ to be~\cite{AlaviHarati86}
\begin{equation}
g_{M1}(s) = \frac{a_1}{m^2_{\rho}-s} + a_2
\end{equation}
with $a_2 = - 1.35$ and $\frac{a_1}{a_2} = - 0.737 \, GeV^2$. The pion charge form factor is adequately
represented by
\begin{equation}
f_{\pi}^{em}(s) = \frac{1}{1-\frac{s}{m^2_{\rho}}}
\end{equation}

\begin{figure}
\center
\makebox[6.5cm]{
\resizebox{6.5cm}{4.9cm}
{\includegraphics*{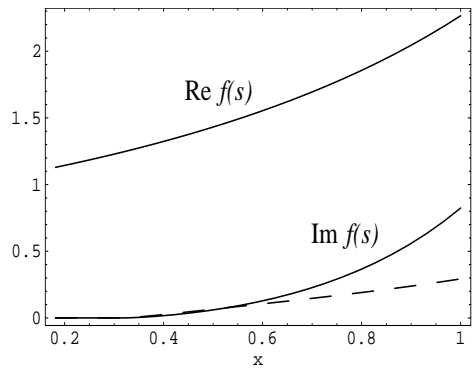}}
}
\caption{Comparison of ${\rm Im}f(s)$ calculated in this paper, with ${\rm Re} f(s)$ as given
by BMS parametrization with $\alpha_{K^*} = -0.19$. Dotted line indicates ${\rm Im} f(s)$ as
obtained from a complex $\rho$-pole
$\left( 1 - \frac{s}{m_{\rho}^2}-i\frac{\Gamma_{\rho}}{m_{\rho}} v^3 \right)^{-1}$.
\label{ReImfs}}
\end{figure}

In terms of $g_{M1}(s)$ and $f_{\pi}^{em}(s)$, we have calculated the imaginary part of the
$K_L \gamma^* \gamma$ form factor to be
\begin{equation}
{\rm Im} f(s) = \left[ \frac{\alpha}{6} \frac{|f_S|}{m_K} \frac{s}{m_K^2} \frac{1}{f_{K \gamma \gamma}(0)}
 \right] f_{\pi}^{em}(s) g_{M1}(s) v^3 \Theta (s-4 m_{\pi}^2)
\end{equation}
with $v = \sqrt{1- \frac{s}{m^2_K}}$, $s$ being the mass of the virtual photon. This is plotted in
Fig.~\ref{ReImfs}, where we also show the real part of $f(s)$, taken to be the BMS form factor.
For comparison, we also indicate in Fig.~\ref{ReImfs} the result for ${\rm Im} f(s)$
that one would obtain from postulating a complex $\rho$-pole with a momentum-dependent decay width,
$f(s) \approx \left( 1 - \frac{s}{m_{\rho}^2} - i \frac{\Gamma_{\rho}}{m_{\rho}} v^3 \right)^{-1}$.

We are now in a position to calculate the decay spectrum of $K_L \to \mu^+ \mu^- \gamma$, taking into
account the interference of conversion and bremsstrahlung. Defining
$x_{\gamma} = \frac{2 E_{\gamma}}{m_K} = (1 - \frac{s}{m_K^2})$ as the scaled photon energy
($0 < x_{\gamma} < 1 - 4 r$, $r=\frac{m^2_l}{m^2_K}$), the distribution in $x_{\gamma}$ is given by
\begin{equation}
\frac{d \Gamma (K_L \to \mu^+ \mu^- \gamma )}{d x_{\gamma}} = \left( \frac{1}{12}
\frac{\alpha \pi}{(2 \pi)^3} m^3_K \right) \left[ Dal+ Int + Brem \right]
\end{equation}
where the three terms, denoting Dalitz, interference and bremsstrahlung contributions, are given by
\begin{eqnarray}
Dal & = & | f_{K \gamma \gamma}(0) |^2 |f(s)|^2 \frac{m_K^2}{s^2} x^2_{\gamma} v ( 1 - x_{\gamma} + 2r)
\nonumber \\
Int & = & f_{K \gamma \gamma} (0) {\rm Re} \left( f_{Kll}^* f(s) \right) \frac{1}{s} \frac{m_{\mu}}{m_K} x_{\gamma}^2
\ln \frac{1 + v}{1- v} \nonumber \\
Brem & = & - |f_{Kll}|^2 \frac{1}{m_K^2} \left[ 2 v \frac{1- x_{\gamma}}{x_{\gamma}} + \left( 2 +
4 \frac{r}{x_{\gamma}} - \frac{2}{x_{\gamma}} - x_{\gamma} \right) \ln \frac{1 + v}{1-v}\right]
\end{eqnarray}

\begin{figure}
\center
\makebox[6.5cm]{
\resizebox{6.5cm}{4.9cm}
{\includegraphics*{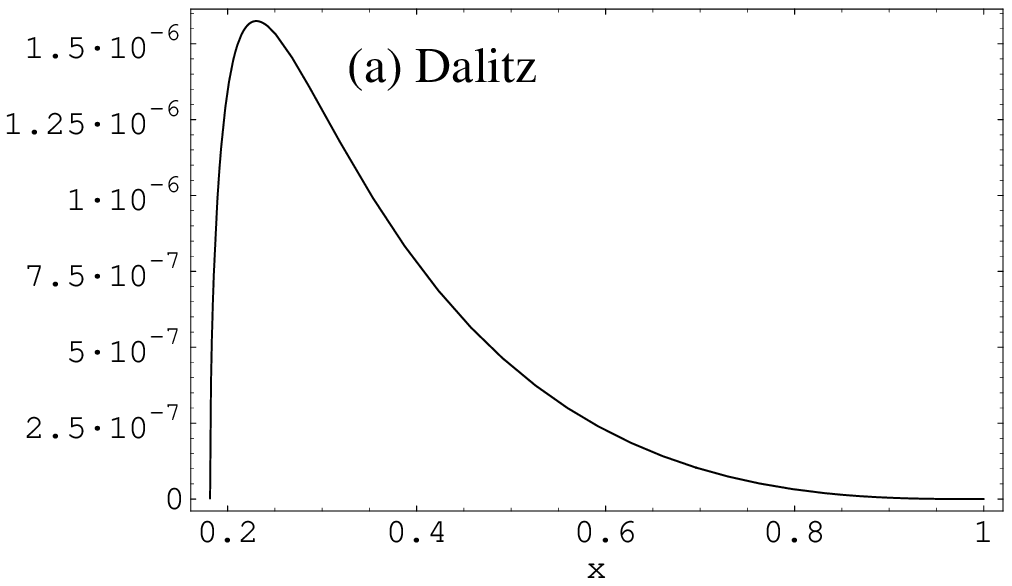}}
}
\makebox[6.5cm]{
\resizebox{6.5cm}{4.9cm}
{\includegraphics*{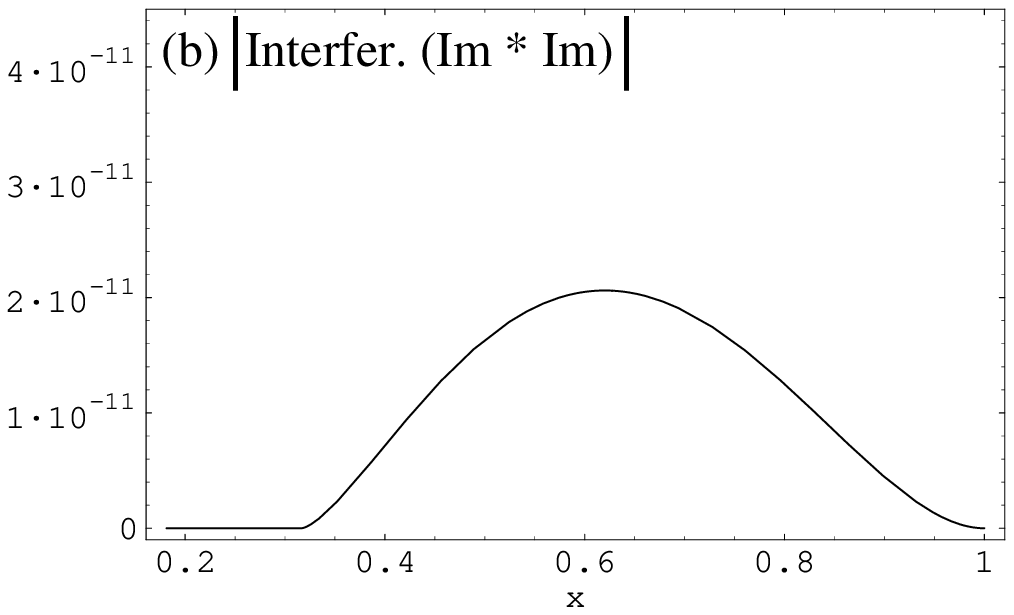}}
}
\makebox[6.5cm]{
\resizebox{6.5cm}{4.9cm}
{\includegraphics*{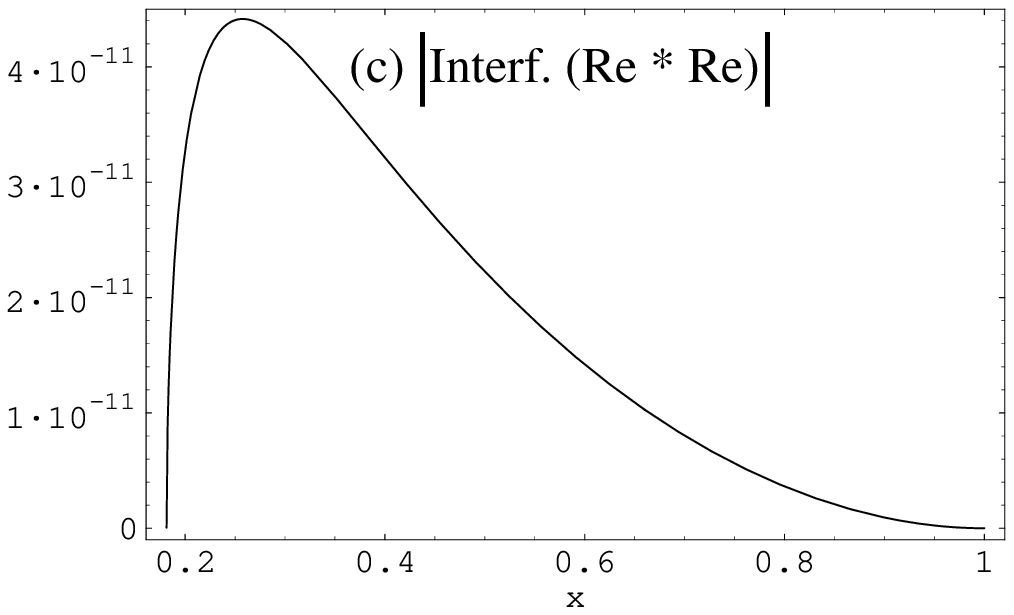}}
}
\makebox[6.5cm]{
\resizebox{6.5cm}{4.9cm}
{\includegraphics*{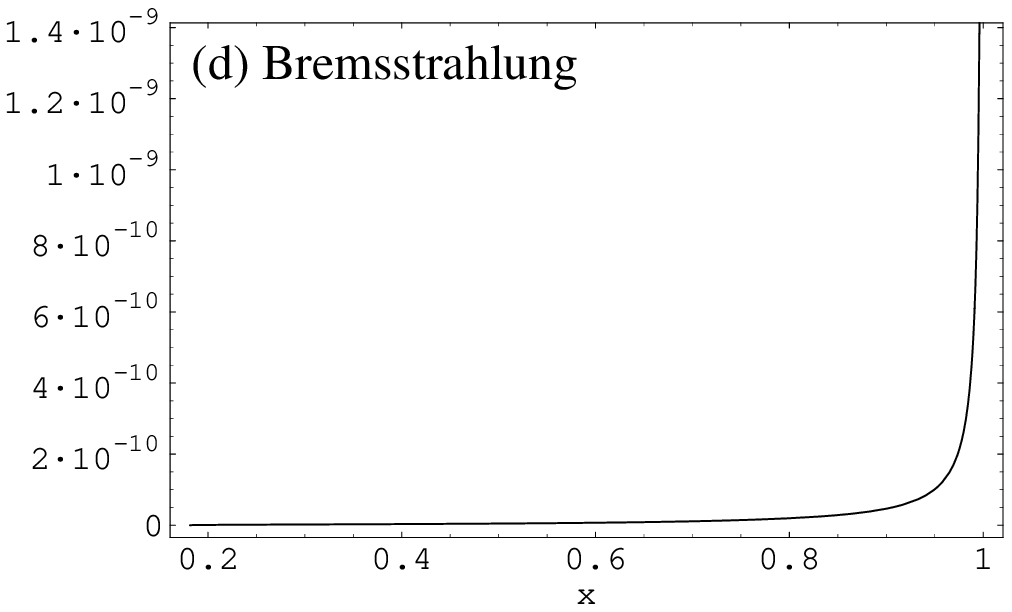}}
}
\caption{Contributions to $d Br \left( K_L \to \mu^+ \mu^- \gamma \right) / dx$ from
(a) Dalitz-pair spectrum, (b) interference of bremsstrahlung with Dalitz amplitude (imaginary parts),
(c) interference between real parts, (d) pure bremsstrahlung. (Interference terms shown in
modulus only.)
\label{AllInSmallPieces}}
\end{figure}

The different contributions to the decay rate $\frac{d \Gamma (K_L \to \mu^+ \mu^- \gamma)}{d x}$
($x \equiv 1 - x_{\gamma} = \frac{s}{m^2_K}$) are plotted in Fig.~\ref{AllInSmallPieces}. The Dalitz pair spectrum shown in
fig.~\ref{AllInSmallPieces}a dominates up to $x \approx 0.95$, and essentially accounts for the measured
branching ratio $Br(K_L \to \mu^+ \mu^- \gamma) \sim 3.7 \times 10^{-7}$. The interference terms proportional to
${\rm Re} f_{Kll} \, {\rm Re} f(s)$ and ${\rm Im} f_{Kll} \, {\rm Im} f(s)$ are shown separately in
Fig.~\ref{AllInSmallPieces}b and \ref{AllInSmallPieces}c, where we have chosen $|{\rm Re} f_{Kll}|$ equal
to the maximum value allowed by the experimental rate of $K_L \to \mu^+ \mu^-$ (Eq.~(\ref{KTeVMeasure})).
The interference terms are competitive with the Dalitz pair spectrum only in the region $x \geq 0.97$,
where the branching ratio is of order $10^{-11}$ per unit of $x$. In the same region the pure
bremsstrahlung contribution shown in Fig.~\ref{AllInSmallPieces}d begins to dominate the spectrum.

\begin{figure}
\center
\makebox[6.5cm]{
\resizebox{6.5cm}{4.9cm}
{\includegraphics*{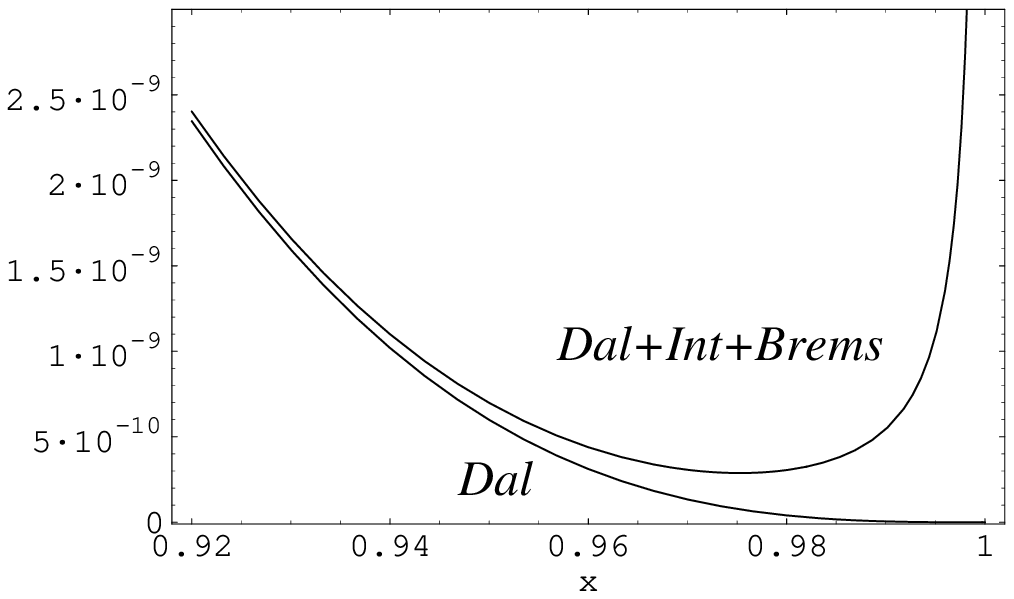}}
}
\caption{Shape of $\mu^+ \mu^-$ invariant mass spectrum in the end-point region (full spectrum
compared to Dalitz-pair contribution).\label{TheWholePicture}}
\end{figure}

The principal conclusion of this paper is embodied in Fig.~\ref{TheWholePicture}, which shows the full
spectrum $\frac{d Br(K_L \to \mu^+ \mu^- \gamma)}{dx}$ in the interesting end-point interval
$0.92 < x < 1.0$. In this region, the spectrum deviates in a systematic way from the monotonically
decreasing Dalitz pair spectrum, going through a minimum around $x \approx 0.97$, and then rising sharply
as $x$ approaches unity, in the manner characteristic of bremsstrahlung. The interference terms affect the
detailed shape of the spectrum in the neighbourhood of the minimum, but only at a level $\approx 10^{-11}$
in the differential branching ratio. In particular, the difference between choosing
${\rm Re} f_{Kll} / {\rm Im} f_{Kll}$ to be $+0.23$ or $-0.23$ is almost unobservable with the resolution
chosen in Fig.~\ref{AllInSmallPieces}.

We conclude that while the spectrum of the decay $K_L \to \mu^+ \mu^- \gamma$ should show an interesting
departure from the Dalitz pair spectrum for large invariant masses, there is little realistic prospect of
being able to extract the real part of the $K_L \to \mu^+ \mu^-$ amplitude from such a measurement. As
a by-product of our analysis, we have determined the imaginary part of the $K_L \to \gamma^* \gamma$ form
factor, relying entirely on an empirical measurement of the spectrum of $K_L \to \pi^+ \pi^- \gamma$.
This imaginary part provides a well-defined correction to existing models of this form factor.

Acknowledgement:
One of us (P.P.) wishes to thank the Alexander von Humboldt Foundation for the award of a post-doctoral
fellowship, and the Institute of Theoretical Physics E, RWTH Aachen, for their hospitality.

\begin{thebibliography}{99}
\bibitem{Sehgal73} L.~M.~Sehgal, Phys.~Rev.~D {\bf 7}, 3303 (1973); and references therein.
\bibitem{Bergstroemetal} L.~Bergstr\"om, E.~Mass\'o and P.~Singer, Phys.~Lett.~B {\bf 249}, 141 (1990).
\bibitem{Fanti;LaDue} V.~Fanti {\it et al} (NA48 Collaboration), Phys.~Lett.~B {\bf 458}, 553 (1999);
J.~LaDue (KTeV Collaboration), talk at DPF 2002, May 27, 2002.
\bibitem{AlaviHarati;Fanti} A.~Alavi-Harati {\it et al} (KTeV Collaboration), Phys.~Rev.~Lett. {\bf 87},
071801; V.~Fanti {\it et al} (NA48 Collaboration), Z.~Phys.~C {\bf 76}, 653 (1997).
\bibitem{BreeseQuinn} G.~Breese Quinn, Ph.D. dissertation, University of Chicago, June 2000.
\bibitem{DAmbrosioetal} G.~D'Ambrosio, G.~Isidori and J.~Portel\'es, Phys.~Lett.~B {\bf 423}, 385 (1998).
\bibitem{Sehgal69} L.~M.~Sehgal, Phys.~Rev. {\bf 183}, 1511 (1969).
\bibitem{AlaviHarati87} A.~Alavi-Harati {\it et al} (KTeV Collaboration), Phys.~Rev.~Lett. {\bf 87}, 071801.
\bibitem{Sehgal;Leusen} L.~M.~Sehgal and J.~van~Leusen, Phys.~Rev.~Lett. {\bf 83}, 4933 (1999).
\bibitem{AlaviHarati86} A.~Alavi-Harati {\it et al} (KTeV Collaboration), Phys.~Rev.~Lett. {\bf 86}, 761 (2001).
\end{thebibliography}
\end{document}